 \documentclass[12pt]{article}

\usepackage{graphicx}
\usepackage{epsfig}
\usepackage{amsfonts}
 
 \usepackage{amssymb}
\usepackage{enumitem}

\newcommand{\insertplot}[5]{\begin{figure}
 \hfill\hbox to 0.05in{\vbox to #5in{\vfill
 \inputplot{#1}{#4}{#5}}\hfill}
 \hfill\vspace{-.1in}
 \caption{#2}\label{#3}
 \end{figure}}
 \newcommand{\inputplot}[3]{
 \special{ps: plotfile #1}
}
\newcounter{fig}

\def\eqref#1{Eq.~(\ref{#1})}

\newcommand{\ee}{\end{equation}}
\newcommand{\eea}{\end{eqnarray}}
\newcommand{\be}{\begin{equation}}
\newcommand{\bea}{\begin{eqnarray}}

\def\X5sp{{\rm X}_5}
\def\Y3sp{{\rm Y}_3}
\def\Z3sp{{\rm Z}_3}

\begin{document}

\title{  
Phases of rotating black objects 
\\
in $d=5$ Einstein-Gauss-Bonnet theory
}


\author{
{\large Burkhard Kleihaus$^1$, Jutta Kunz$^1$}  
and
{\large Eugen Radu$^2$} 
\\ 
\\
{\small $^1$  Institute of Physics, University of Oldenburg, Postfach 2503, D-26111 Oldenburg,
Germany}  
\\ 
{\small $^2$  Centre for Research and Development  in Mathematics and Applications (CIDMA)},
\\
{\small
Campus de Santiago, 3810-183 Aveiro, Portugal}
}

\maketitle

\abstract{
We consider several 
different classes of asymptotically flat, rotating  black objects in 
$d=5$ Einstein-Gauss-Bonnet (EGB) theory.
These are first the black holes with 
two equal-magnitude angular momenta,
in which case extremal configurations are studied as well.
Numerical evidence is also
given for the existence of EGB
generalizations of the Myers-Perry black holes with a single plane of rotation 
and of the Emparan-Reall balanced black rings.
All solutions approach asymptotically the Minkowski background and
 present no singularities outside 
and on the horizon.
The numerical results suggest that for 
any mass of the solutions and any topology of the horizon, 
the rotating configurations  exist up to a maximal value 
of the GB coupling constant,
 while the solutions with a spherical horizon topology still 
satisfy the Einstein gravity bound on angular momentum.
}


\section{Introduction}
 
In $d=5$  spacetime dimensions  the
Einstein-Gauss-Bonnet (EGB) model   
provides the most general theory of gravity 
which
includes higher order curvature terms, 
while keeping the equations of motion to second order \cite{Lovelock:1971yv}. 
Apart from being of mathematical interest
and providing a natural generalization of General Relativity (GR),
the  Gauss-Bonnet (GB) term appears 
in the low-energy effective action
for the compactification of M-theory 
on a Calabi-Yau threefold  \cite{Antoniadis:1997eg}
and also enters
the one-loop corrected
effective action of heterotic string theory~\cite{1,Myers:1987yn}.

 The  Black Hole (BH) solutions
of EGB gravity have  been studied by various authors,
starting with the Ref.~\cite{Deser},
where a generalization of the Schwarzschild-Tangherlini BH
\cite{Tangherlini:1963bw}
has been found. 
These solutions possess a variety of 
new features; for example, 
 their entropy includes a GB
contribution 
\cite{Jacobson:1993xs,Wald:1993nt},
with the existence of 
 a branch of small static BHs 
which are thermodynamically stable.

However, the complexity of the EGB theory 
makes the task of finding solutions 
beyond those in \cite{Deser}
 a highly non-trivial problem 
\cite{Garraffo:2008hu,Charmousis:2008kc}. 
In particular, no EGB closed form rotating solutions
are known yet,
 and it was proven in \cite{Anabalon:2009kq} that the
Kerr-Schild ansatz does not work in this model.
Nevertheless, a number of partial results 
(including perturbative exact solutions
\cite{Kim:2007iw,Konoplya:2020fbx}
and numerical
non-perturbative results
for configurations with  symmetry enhancement
 \cite{Brihaye:2008kh,Brihaye:2010wx})
support the idea that
EGB rotating solutions actually exist.

This issue is of special interest, since, 
as discovered by Emparan and Reall \cite{Emparan:2001wn}
rotation allows in this case for Black Ring  (BR) solutions,
in addition to the generalization  of the Kerr BH found by  Myers and Perry \cite{Myers:1986un}. 
This (asymptotically flat,  vacuum GR) solution 
has a horizon with topology $S^2\times S^1$,
while the  MP BH 
has a  horizon topology $S^3$.
This made clear that a number of well known results in $d=4$ 
gravity
stop to be valid in higher dimensional GR.
Therefore it would be interesting to find whether the situation persists
for other models of gravity. 
 
In this work we address the question on how the GB term affects the
phase structure of several different types of $d=5$ rotating black objects.
We shall first consider BHs
with two equal-magnitude angular momenta,
 extending the results found in 
Ref.~\cite{Brihaye:2010wx} by including the set of extremal solutions.
Black objects 
 rotating in a single plane are studied as well, 
and we report EGB generalizations of both MP BHs and rotating  BRs.
All solutions are found within a nonperturbative approach, 
by directly solving the second order field equations
with suitable boundary conditions.
 
\section{The model and the static limit}

\subsection{Action, equations and scaled quantities}

Working in units with $c=G=1$,
we consider the EGB action in five space-time dimensions
\begin{equation}
I = \frac{1}{16\pi  } 
   \int_\mathcal{M}d^5x \sqrt{-g}\left[ R + \alpha L_{\rm GB}\right]
   \ , 
\label{action}
\end{equation}
where   $\alpha$ is the GB coefficient with dimension $(length)^2$. 
In string theory,
the GB coefficient is positive,
and this is the only case considered here.
$R$ denotes the Ricci scalar, and
\begin{equation}
L_{\rm GB}  = R^2 - 4 R_{\mu\nu}R^{\mu\nu} 
            + R_{\mu\nu\rho\sigma}R^{\mu\nu\rho\sigma} 
\label{GBterm}
\end{equation}is 
the GB term, with Ricci tensor $R_{\mu\nu}$ and 
Riemann tensor $R_{\mu\nu\rho\sigma}$.

The variation of the action (\ref{action}) with respect to the
metric tensor yields the EGB equations	    
\begin{equation}
 G_{\mu\nu} +\alpha H_{\mu\nu}=0 \ , 
\label{EGBeqs}
\end{equation}
where
\begin{eqnarray}
G_{\mu\nu} & = & R_{\mu\nu} -\frac{1}{2} g_{\mu\nu} R \ , 
\nonumber \\
H_{\mu\nu} & = & 2\left[R R_{\mu\nu} -2 R_{\mu\rho}R^\rho_\nu
                        -2 R_{\mu\rho\nu\sigma}R^{\rho\sigma}
			+R_{\mu\rho\sigma\lambda}R_\nu^{\ \rho\sigma\lambda}
		   \right]
		   -\frac{1}{2}g_{\mu\nu}L_{\rm GB}	.
\nonumber 
\end{eqnarray}


The solutions discussed in this work
approach asymptotically 
the $d=5$ Minkowski spacetime background,
with a line element
\begin{eqnarray}
\label{mink}
ds^2=dr^2+r^2 d \Omega_3^2-dt^2,
~~{\rm with}~~
d\Omega_3^2=d\theta^2+\sin^2\theta d\varphi_1^2+\cos^2\theta d\varphi_2^2,~~
\end{eqnarray} 
where 
$\theta  \in [0,\pi/2]$, $(\varphi_1,\varphi_2) \in [0,2\pi]$,
while $r$ and $t$ denote the radial and time coordinate, respectively.
Apart from the mass $M$, they possess a nonzero angular momentum $J$
(or two  equal angular momenta, $J_1=J_2=J$),
with $(M,J)$ read as usual from the far field asymptotics of the metric 
functions
$g_{tt}$
and 
$g_{\varphi_i t}$, respectively.
The horizon quantities of main interest
are the Hawking temperature $T_H$,
event horizon area $A_H$,
event horizon velocity 
$\Omega_H$
(with $\Omega_{H(1)}=\Omega_{H(2)}=\Omega_H$
for BHs rotating in two planes),
and also the
entropy 
$S$,
which is the sum of one quarter of the event horizon area
(the Einstein gravity term) plus a GB correction 
 \cite{Wald:1993nt}
\begin{eqnarray}
\label{entropy-gen}
S=\frac{1}{4}\int_{\Sigma_h} d^{3}x \sqrt{  h}(1+2 \alpha R_\Sigma),
\end{eqnarray}
 where $h$ is the determinant of the induced metric on the horizon
and $ R_\Sigma$ is the event horizon curvature.
Also, 
the  solutions satisfy the 1st law of thermodynamics
\begin{eqnarray}
dM=T_H dS+k\Omega_H dJ,
\end{eqnarray}
(with $k=1$ or $k=2$ the number of planes of rotation).

In what follows 
we shall consider several
quantities 
of interest
normalised $w.r.t.$ the mass of the solutions
and  define\footnote{
Various numerical factors 
in eq.~(\ref{scaled1})
have been chosen such that
$t_H=a_H=s=1$ 
in the static limit with $\alpha=0$,
while 
the maximal value for Einstein gravity BH solutions
is 
$j=1$.
}
 \begin{eqnarray}
\label{scaled1} 
a_H=\frac{3}{32}\sqrt{\frac{3}{2 \pi}}\frac{A_H}{M^{3/2}},~
s=\frac{3}{8}\sqrt{\frac{3}{2 \pi}}\frac{S}{M^{3/2}},~
t_H=4\sqrt{\frac{2\pi}{3}}T_H \sqrt{M},~
j=\frac{3}{4}\sqrt{\frac{3\pi}{2}}\frac{kJ}{M^{3/2}}.~{~}
\end{eqnarray} 	
The (dimensionless) ratio 
between the parameter $\alpha$
and the mass is also important, 
and we define
\begin{eqnarray}
\label{x}
x=c_0 \frac{\alpha}{ M}, ~~~{\rm with}~~c_0=\frac{3\pi }{4} .
\end{eqnarray}

\subsection{The Schwarzschild-Tangherlini solution in EGB theory}
  
The static, spherically symmetric EGB BH solution\footnote{Static EGB
solutions with a
$S^2\times S^1$ horizon topology ($i.e.$ BRs)
are also known to exist \cite{Kleihaus:2009dm}, although not in closed form.
However, these solutions (still) possess a conical singularity, 
and thus are physically less interesting.}  
has a relatively simple form \cite{Deser}
 \begin{eqnarray}
\label{ST-EGB}
ds^2=\frac{dr^2}{N(r)}+r^2 d\Omega_3^2-N(r) dt^2,
~~
{\rm with}~~
N(r)=1+\frac{ r^2}{ 4\alpha}
 \bigg (
  1-\sqrt{1+ \frac{8\alpha(r_h^2+2\alpha)}{r^{4}} }
 \bigg).~{~~}
\end{eqnarray}
The parameter
$r_h>0$ denotes the event horizon radius, 
with $N(r)=\frac{2r_h}{r_h^2+4\alpha}(r-r_h)+\dots$,
as $r\to r_h$.
While $r_H$ can be arbitrarily large, 
 the limit $r_h\to 0$ 
is nontrivial,  
with no horizon and
$
N(r)=1- {\alpha}/(r^2({1+\sqrt{1+\frac{\alpha^2}{r^4}}}))~
$
a strictly positive function.
However,
 this configuration is pathological,  
$r=0$ corresponding to a naked singularity,
with a diverging Ricci scalar.

The expressions of various quantities of interest for the spherically symmetric BH
solutions
are
\begin{eqnarray}
\label{SGB-TH}
 M=\frac{3 \pi}{8 }(r_h^2+2\alpha),~~T_H=\frac{r_h}{2\pi(r_h^2+4\alpha)},~~
A_H=2\pi^2 r_h^3,~~
 S=\frac{\pi^2 r_h^3}{2}(1+\frac{12\alpha}{r_h^2}).
\end{eqnarray}  
As such, the mass spectrum of these EGB BHs 
is bounded from below by the mass corresponding to the nakedly singular configuration,
$ M>{3 \pi \alpha}/{4 }$,
a result which is found to hold also for spinning generalizations.

A straightforward computation leads to the following
expression of several scaled quantities, cf.~(\ref{scaled1})
\begin{eqnarray}
a_H= (1-x)^{3/2},~~s=\sqrt{1-x}(1+5x),~~t_H=\frac{\sqrt{1-x}}{1+x},
~~~{\rm with }~~~x= \frac{3\pi\alpha}{ 4M},
\end{eqnarray}  
where $0 \leq x \leq 1$.
The limit $x=0$
corresponds to the Schwarzschild-Tangherlini BH in pure Einstein gravity.
As $x\to 1$, the minimal mass
(nakedly singular) solution is approached, with the (scaled) horizon size,
entropy and temperature going to zero. 
It is interesting to remark that
 the scaled entropy varies between zero and
a maximal value
\begin{eqnarray}
 s_{max}=4\sqrt{2/5}\simeq 2.52892,
\end{eqnarray} 
which is  approached for a special configuration with $x=3/5$
  (marked with a black dot in Fig.~\ref{fig1}, left panel (middle)).
Therefore, for given $\alpha$ and 
a range of $s$, there are two different solutions with the same entropy.
At the same time, the scaled horizon area  and temperature 
varies monotonically between one and zero,
see the corresponding curves 
in  Fig.~\ref{fig1}.


\section{Rotating black holes: the case of equal angular momenta }

\subsection{The ansatz and particular cases}

For these solutions, the isometry group is enhanced from $\mathbb{R}_t \times U(1)^{2}$
to $\mathbb{R}_t \times U(2)$ (where $\mathbb{R}_t$ denotes the time translation),
a symmetry enhancement which allows to factorize the angular dependence of the metric. 
The line element    
takes a simple form in terms of
the left-invariant
one-forms $\sigma_i$ on $S^3$, with
\begin{eqnarray}
\label{metric2}
ds^2 = f_1(r) dr^2
  + \frac{1}{4}f_2(r)(\sigma_1^2+\sigma_2^2)+\frac{1}{4}f_3(r) \big(\sigma_3-2w(r) dt \big)^2
-f_0(r) dt^2,
\end{eqnarray}
where
$\sigma_1=\cos \psi d\bar \theta+\sin\psi \sin  \theta d \phi$,
$\sigma_2=-\cos \psi d\bar \theta+\cos\psi \sin  \theta d \phi$,
$\sigma_3=d\psi  + \cos  \theta d \phi$
and 
we define
$2\theta=\bar \theta$,
 $\varphi_1-\varphi_2=\phi$,
  $\varphi_1+\varphi_2=\psi$
	(with $\{\theta,\varphi_1,\varphi_2 \}$
	the angular coordinates in (\ref{mink})).
	This geometry describes a fibration of $AdS_2$ 
	over the homogeneously squashed $S^3$ with symmetry
group $SO(2, 1)\times SU(2)\times U(1)$.
The horizon is located at some $r=r_h$ (where $f_0(r_h)=0$), with the 
induced horizon metric
\begin{eqnarray}
\label{metric2-hm}
d\Sigma^2=h_{ij}dx^idx^j =  \frac{1}{4}\left(
 f_2(r_h)(\sigma_1^2+\sigma_2^2) + f_3(r_h)  \sigma_3^2
\right).
\end{eqnarray}
 
\medskip

For a given mass, the ($\alpha =0$) MP BHs
exhibit a 
similar behaviour to that found for $d=4$ Kerr BHs,
forming a one parameter family of  solutions 
which 
interpolates between the static limit
($j=0$)
and an extremal configuration with $j=1$, $t_H=0$ 
 and a nonzero horizon area\footnote{The 
$\alpha=0$ MP solution can be written in the form (\ref{metric2}),
with the
expression of the functions 
$f_i(r)$
and
$w(r)$
given  $e.g.$ 
in Section 2.3 of
Ref.~\cite{Brihaye:2010wx}.
Also, this solution has 
$a_H=s=\frac{1}{2} 
\left(
1+\sqrt{1-j^2}
\right),$
and
$
t_H= {2 \sqrt{1-j^2}}/{(1+\sqrt{1-j^2})}$.
}.
As expected, all these BHs possess 
generalizations with $\alpha \neq 0$,
a study of the non-extremal  solutions
being reported in Ref.~\cite{Brihaye:2010wx}.
Most of the work there has been performed for 
a metric gauge choice with
$f_2(r)=r^2$,
the  
metric functions $f_0(r),f_1(r),f_3(r)$ 
and $w(r)$ 
being found numerically as solutions of a complicated set of ordinary differential equations
with suitable boundary conditions.
A detailed study of these aspects has been reported in 
Ref.~\cite{Brihaye:2010wx} and we shall not repeat them here.

\medskip

Following the same procedure, we have extended 
the results in Ref.~\cite{Brihaye:2010wx},
attempting to obtain a complete scan of the domain of existence 
of the solutions
(in particular, for the region with  $x>0.5$,
as defined in eq.~(\ref{x}),
which was poorly covered in \cite{Brihaye:2010wx}).

Also, 
the results in  Ref.~\cite{Brihaye:2010wx} 
strongly suggest 
that the families of rotating EGB BHs
terminate at extremal configurations. 
Although this special set of solutions
was not constructed in \cite{Brihaye:2010wx},
 the extrapolated results indicated 
that all relevant quantities remain finite 
in the extremal limit, 
while the Hawking temperature vanishes.  

This is indeed confirmed by the results below,
which are found 
by extending the methods  
in \cite{Brihaye:2010wx},
and  constructing directly
the extremal BHs in EGB theory.
These solutions are found for
a form of
 the metric ansatz (\ref{metric2}),
with
$f_1(r)=e^{2F_1(r)}/B_1(r)$,
$f_2(r)=e^{2F_2(r)}u(r)$,
$f_3(r)=e^{2( F_2(r)+F_3(r) ) }u(r)B_3(r)$,
$f_0(r)=e^{2F_1(r)} B_2(r)$,
$w(r)=w_0(r)+W(r)$,
a parametrization
which contains four unknown functions
$F_1(r)$,
$F_2(r)$,
$F_3(r)$
and
$W$,
as well as the
background functions
\begin{eqnarray}
\nonumber
B_1(r)=\frac{r^2 }{u(r) },~
B_2(r)=\frac{r^4 }{u(r)^2+a^4},~
B_3(r)=1+\frac{ a^4 }{u(r)^2},~
w_0(r)=\frac{\sqrt{2} a^3}{u(r)^2+a^4},
\end{eqnarray} 
with 
$u(r)=r^2+a^2$
and 
$a>0$  an input parameter\footnote{The limit 
$F_0=F_1=F_2=W=0$ corresponds 
to the
extremal MP solution in Einstein gravity.}.

In this approach, 
the extremal horizon is located at $r=0$, 
where one can construct an approximate form of the solutions as 
a power series  
in $r$.
A similar approximate solution can be found for large $r$
(with $e.g.$
$F_1= {c_t}/{r^2}+\dots$
and
$W= {c_w}/{r^4}+\dots$),
which reveals the existence of two free constants $c_t$
and $c_w$.
The solutions which smoothly interpolate
between these
asymptotics are found
by 
using similar methods to those described in \cite{Brihaye:2010wx},
and
by solving  numerically
the equations for $(F_i,W)$
with suitable boundary conditions.
The quantities of interest are computed from 
the numerical output, with  
\begin{eqnarray}
&&
M=\frac{3\pi}{4}(a^2-c_t),~~
J=\frac{\pi}{4}(c_w+\sqrt{2}a^3),~
A_H=4\pi^2 \frac{a^3}{\sqrt{2}}e^{3F_2(0)+F_3(0)},~~
\\
\nonumber
&&
 S=\pi^2 \frac{a^3}{\sqrt{2}}e^{3F_2(0)+F_3(0)}
 +4  \pi^2 \sqrt{2}  \alpha  a 
e^{F_2(0)+F_3(0)}  (2-e^{2F_3(0)})~.
\end{eqnarray}

\subsection{The domain of existence and attractors}

\begin{figure}[p!]
\begin{center}
\includegraphics[height=.235\textheight, angle =0]{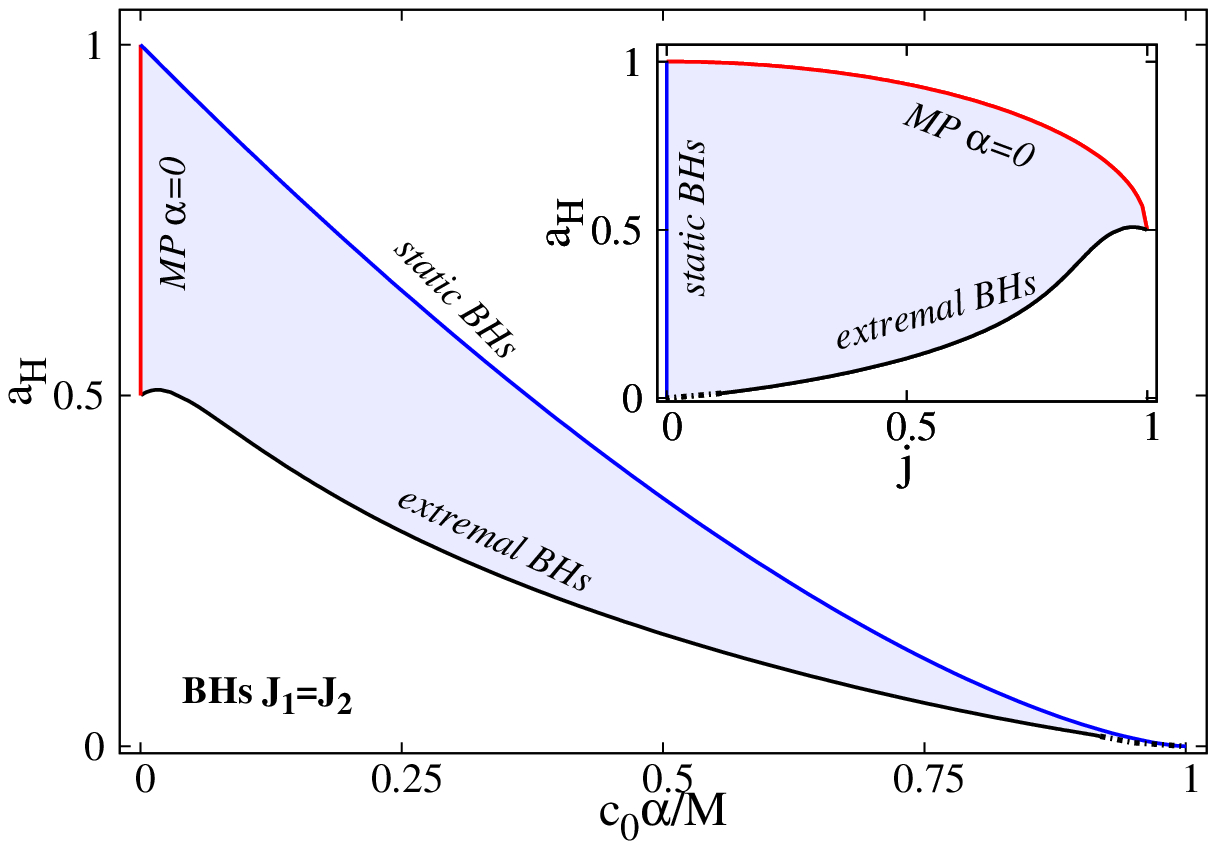} 
\includegraphics[height=.235\textheight, angle =0]{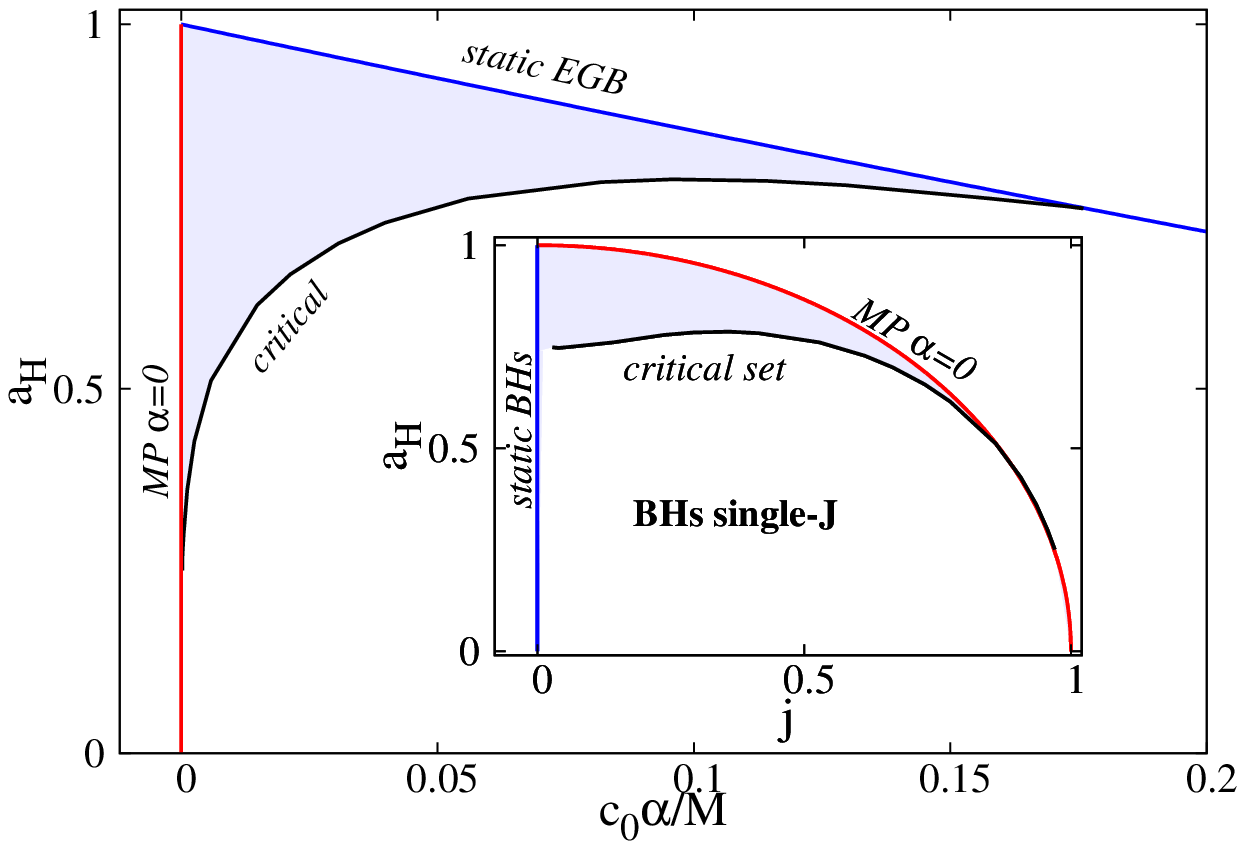} \ \ 
\includegraphics[height=.235\textheight, angle =0]{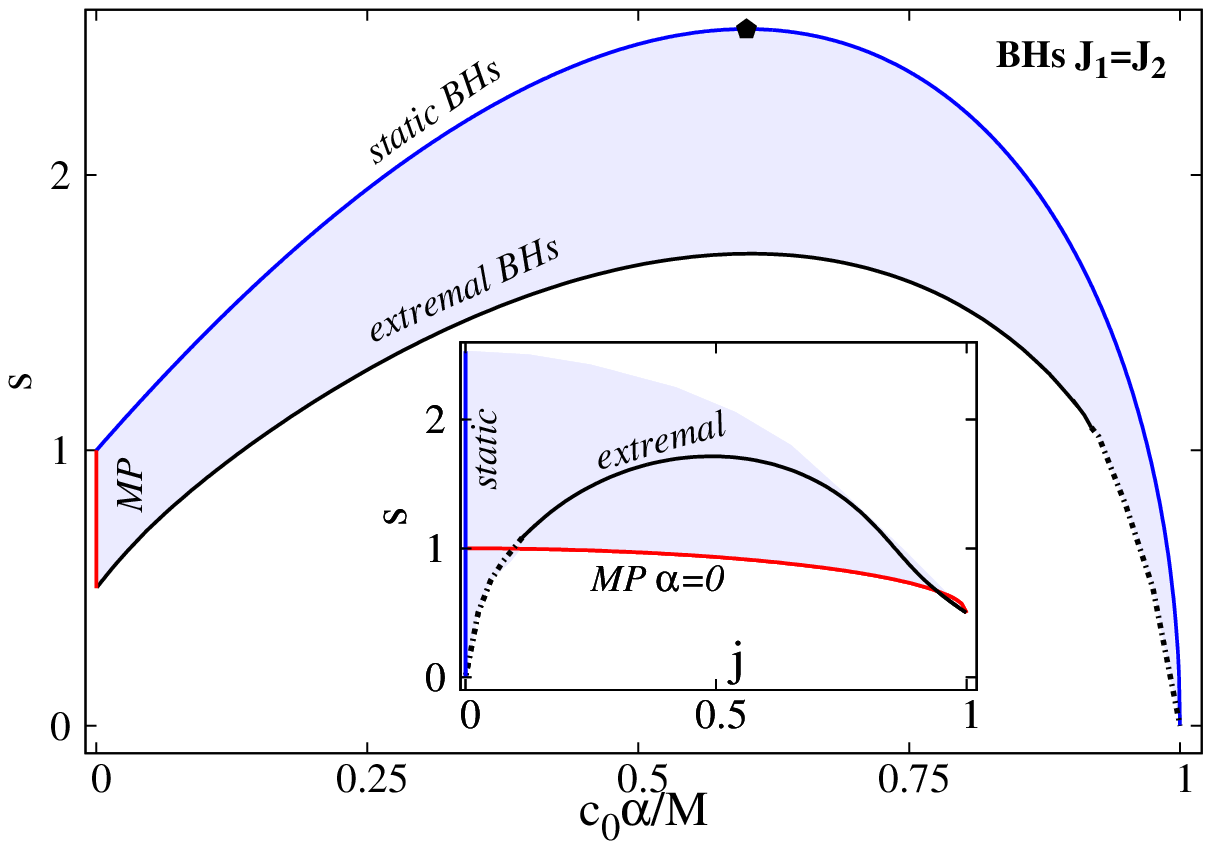} 
\includegraphics[height=.235\textheight, angle =0]{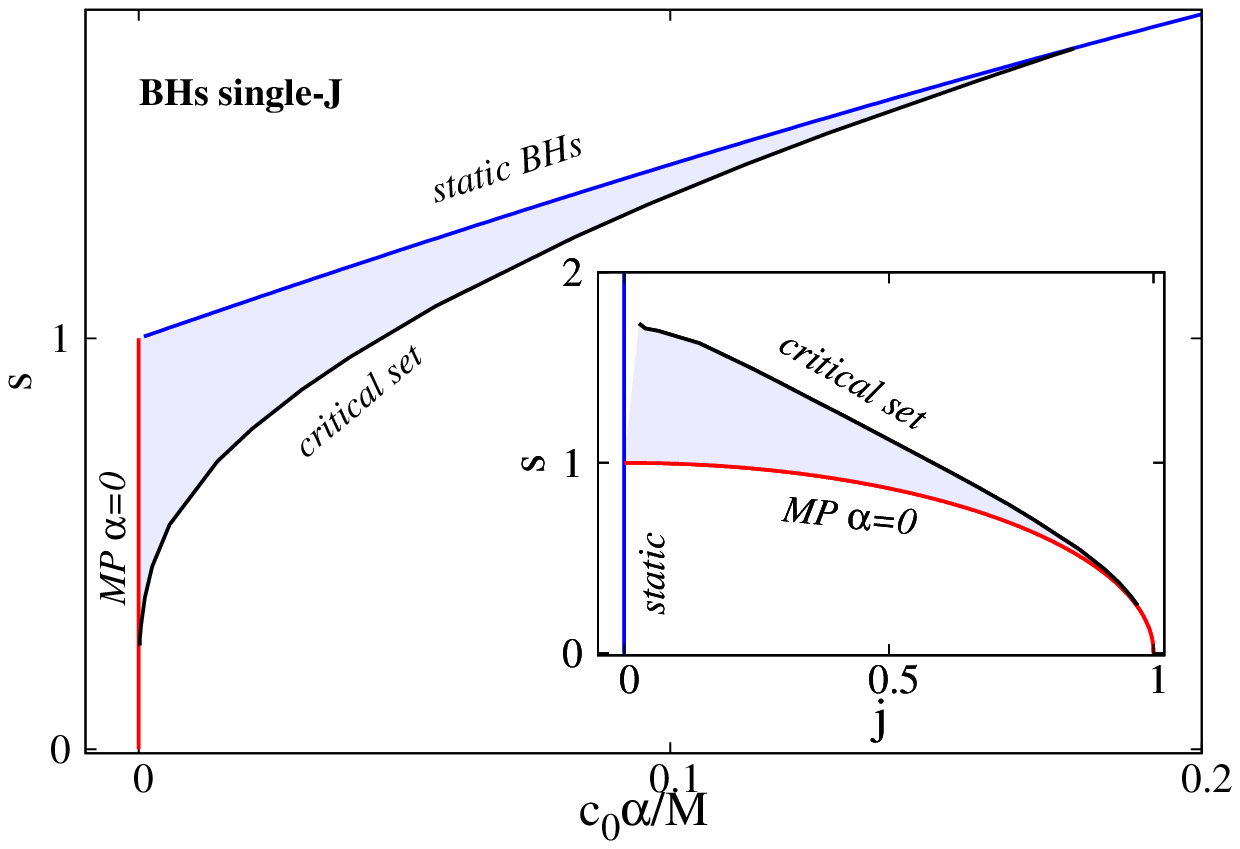} \ \
\includegraphics[height=.235\textheight, angle =0]{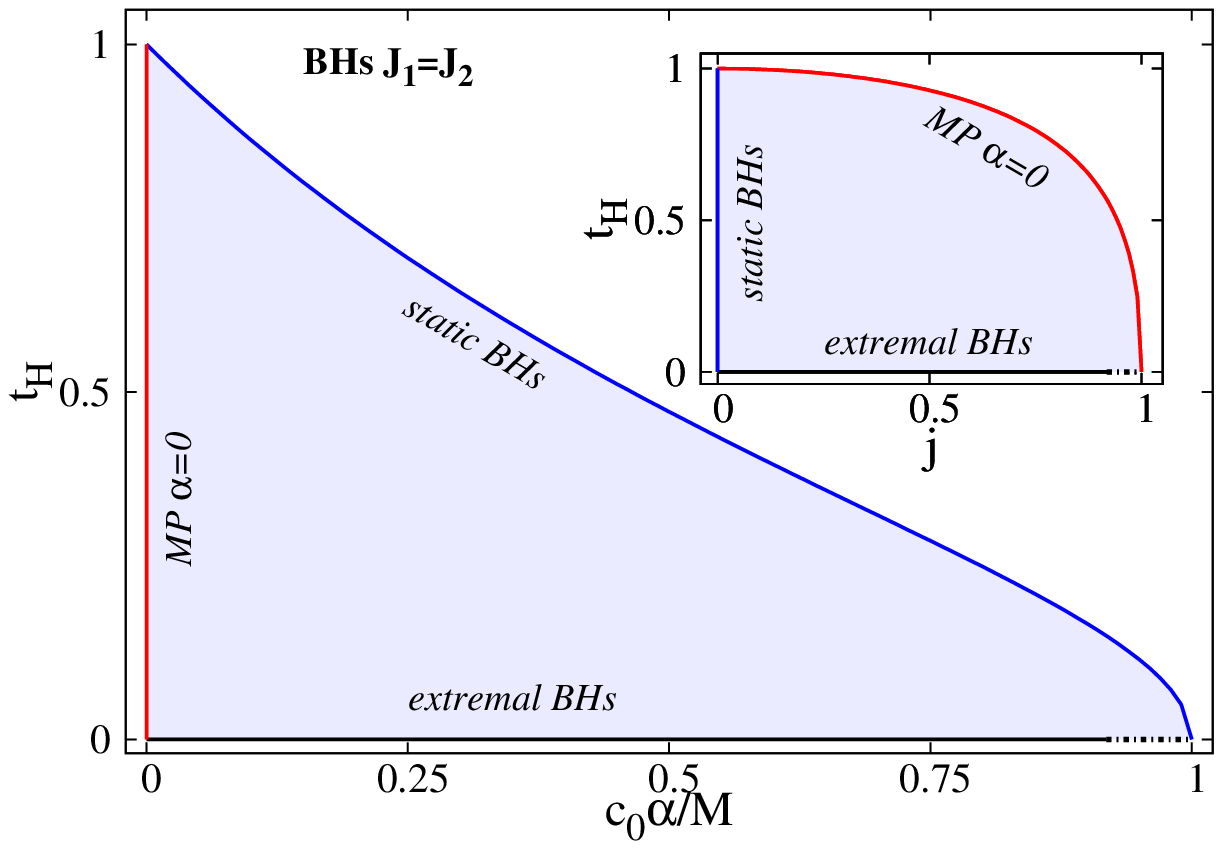} 
\includegraphics[height=.235\textheight, angle =0]{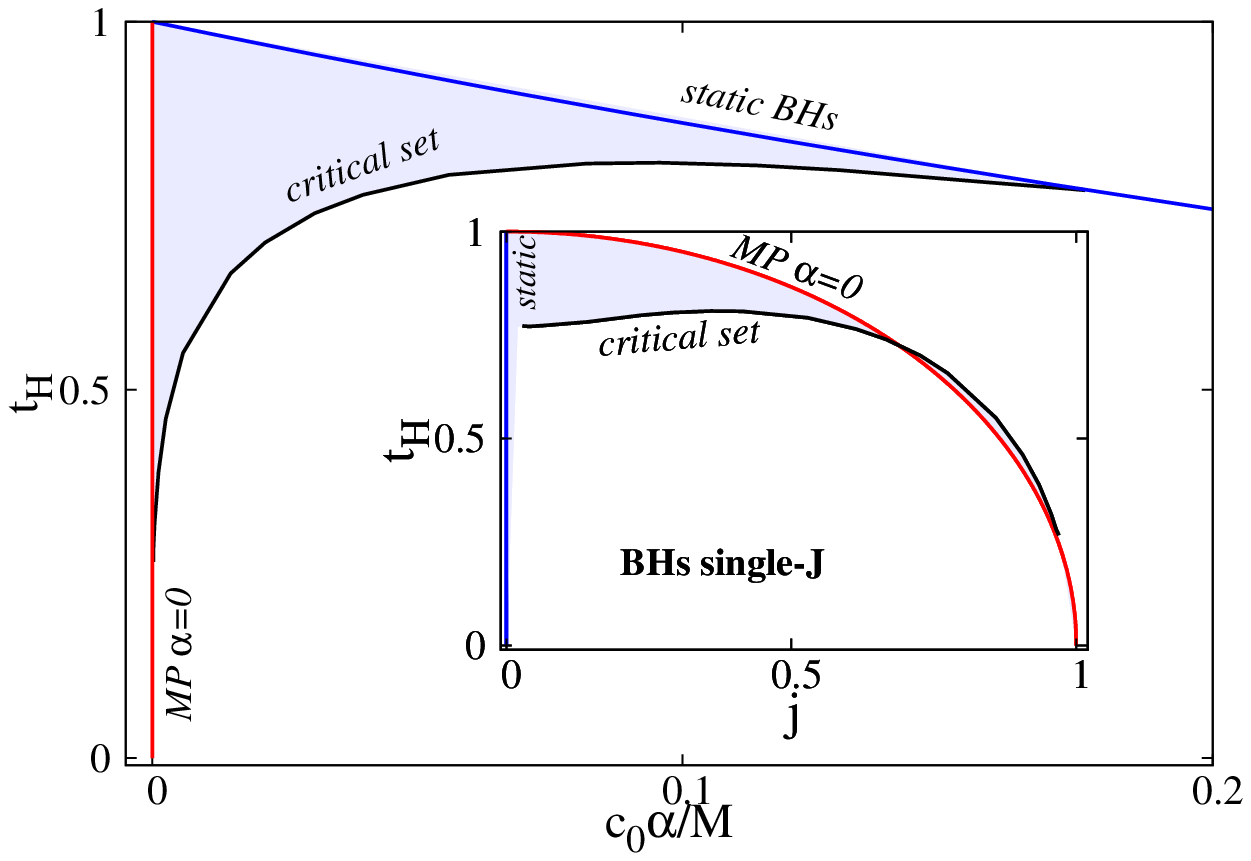} \ \  
\end{center}
  \vspace{-0.5cm}
\caption{ 
{\it Left  panels:} 
The domain of existence of the horizon area, entropy and temperature  
is shown $vs.$~
GB parameter
$\alpha$   
and   $vs.$~$J$ (insets) for EGB black holes with two equal angular momenta.  
{\it Right  panels:} 
The investigated region of the parameter space is shown for 
EGB black holes with a single plane of rotation.   
All quantities are normalized $w.r.t.$ the mass of the solutions,
while $c_0={3\pi }/{4}$. 
}
\label{fig1}
\end{figure}

In Fig.~\ref{fig1} 
(left panels)
we plot the domain of existence of solutions (shaded blue region),
as resulting from 
the extrapolation of around one thousand of data points into the continuum. 
The figure shows that this region
is delimited by\footnote{Note, that  a part of the boundary of the $(j,s)$-domain consists of a configurations with maximal entropy, which do not coincide with the other sets of limiting solutions.}: 
$i)$ the set of static BHs 
discussed in Section 2.2 (blue curve);
$ii)$ the set of extremal BHs (black curve), and
$iii)$ the set of $\alpha=0$ GR solutions corresponding to the MP BHs (red curve).
As one can see, 
the inequality $j\leq 1$
(which is
satisfied by the $\alpha=0$ BHs) 
still holds in the EGB theory.
Moreover,
 the upper bound found for static BHs
$\alpha <4M/(3\pi)$
is still valid for spinning solutions.

Figure \ref{fig1} (left)
includes also the sets of extremal solutions discussed above
(which also emerge as limits of the configurations
 in Ref.~\cite{Brihaye:2010wx}).
As $\alpha$ is varied for a given mass,
these configurations connect the extremal MP limit with a  critical 
configuration.
The  limit is difficult to approach, since
the integration of the equations is becoming increasingly difficult.
Nevertheless, we conjecture that  
this limit of  the extremal set of solutions
corresponds to the  static  singular 
solution discussed in Section 2, with
$
M=\frac{3 \pi}{4 } \alpha,
$
and 
$J=A_H=S=0$.
As such,  the corresponding curves for extremal solutions 
in Fig.~\ref{fig1} (left)
have been extrapolated to this point (dotted black line).  
 
Apart from the numerical results, another indication
supporting this conjecture 
(together with
several analytical results) 
comes from 
the study of an exact EGB solution
describing a rotating squashed $AdS_2\times S^3$ spacetime,
which
corresponds to the neighborhood of the event horizon of an
extremal BH. 
The corresponding metric Ansatz 
is given again by  (\ref{metric2}), with
$f_0=v_1r^2$,
$f_1=v_1/r^2$,
$f_2=v_2$,
$f_3=v_2v_3$
and
$w=-k r$,
 the constant parameters $v_1,v_2,v_3$
	and $k$ 
	being found by solving the EGB equations.
This results in a single parameter family of solutions \cite{Brihaye:2010wx}, 
	which   
takes a relatively simple form in terms of $v_3$ 
(which measures the relative squashing of the $S^3$-sector in (\ref{metric2}), 
 with $0\leq v_3\leq 2$):
 \begin{eqnarray}
\label{at7} 
v_1= \frac{(v_2-4\alpha (3v_3-4)  )(3v_2+4   \alpha(4-v_3))}
{2(4-v3)(3v_2+4 \alpha(8-6v_3)) },
~~
k=(4-v_3)\sqrt{v_2 v_3}\frac{\pi v_1}{2J}~,
\end{eqnarray}
with 
 \begin{eqnarray} 
\label{at8}
v_2=\frac{4\alpha}{v_3-2}
\left(
2v_3^2-7v_3+4
-\sqrt{5v_3^4-34 v_3^3+73 v_3^2-56 v_3+16}
\right).
\end{eqnarray}
%

%
The attractor formalism allows to compute 
the expressions for the angular momentum, event horizon area
and entropy of the solutions, with
%
 \begin{eqnarray}
\label{at9}
&& 
J= \frac{\pi }{4}v_2 v_3
\sqrt{(4-v_3)(v_2+4\alpha (4-3v_3))},~~{~~}
\\
&& 
A_H=2\pi^2 v_2 \sqrt{v_2 v_3},~~
S_{(extremal)}=  \frac{\pi^2}{2 }\sqrt{v_2 v_3}\left(v_2+4\alpha  (4-v_3)\right).
\end{eqnarray}
Therefore the special configuration  
with 
$v_3=0$
 corresponds to the 
critical limiting solution,
which has 
$v_1=\alpha$,
$v_2=0$
and 
$A_H=S=J=0$.

The connection
of the above results
 with the  
extremal BH solutions is straightforward, via
the following identification
  \begin{eqnarray}
v_1=\frac{1}{4}e^{2F_1(0)}a^2,~~
v_2= e^{2F_2(0)}a^2,~~
v_3=2 e^{2F_3(0)},
\end{eqnarray}
and is used 
(together with 
(\ref{at7}), 
(\ref{at8}), 
(\ref{at9}))
to check the accuracy of the numerical results.

\section{Black objects rotating in a single plane: holes and rings}

\subsection{The ansatz and quantities of interest }
	
	The case of  BHs with two equal-magnitude angular momenta
	is rather special, since generically $J_1\neq J_2$.
	However, in the absence of 
	the symmetry enhancement,
	this results in a  
	set of highly nonlinear coupled partial differential equations,
		which are difficult to study.
		In what follows, we shall simplify the problem,
		restricting to configurations with a single plane of rotation.
Two different classes of solutions are considered in this case,
	corresponding to EGB generalizations of (singly spinning) MP BHs
		(with an $S^3$ event horizon topology) 
 and of Emparan-Reall BRs
	(with an $S^2\times S^1$ event horizon topology).

Both types of configurations are constructed within a
metric ansatz\footnote{
 Note that the line element
 (\ref{metric}) 
 can be employed as well in the study of 
 solitonic compact objects, in which case the range of the radial coordinate
is $0\leq r<\infty$.
Such configurations possess no horizon  ($f_0(r,\theta) \neq 0$)
and satisfy a specific set of boundary conditions
at the origin, 
$r=0$ 
(with $f_2=f_3=W=0$ and $\partial_r f_1=\partial_r  f_0=0$),
 while the boundary conditions at $\theta=0,\pi/2$
 and as $r\to \infty$ are similar to those employed for BHs with spherical horizon topology.
Also, one remarks that the static limit of 
the line-element  (\ref{metric})
results in the
Schwarzschild-Tangherlini-EGB solution 
in isotropic coordinates,
$i.e.$ 
with
a different  radial coordinate than in (\ref{ST-EGB}).
}
 with five unknown functions
($f_i$, $w$): 
\begin{eqnarray}
\label{metric}
ds^2= {f_1(r, \theta) }(dr^2+r^2 d\theta^2)
+ f_2(r, \theta) (d\varphi_1- w(r, \theta)dt )^2 
+ f_3(r, \theta) d\varphi_2^2
-f_0(r, \theta) dt^2.
~~{~~}
 \end{eqnarray} 
For both BHs and BRs, the event horizon is localized at constant radius, $r=r_h$,
where $f_0(r_h)=0$.
Expanding the EGB equations in the vicinity of the horizon in powers of $r-r_h$, one finds   
$f_i(r,\theta)=f_{i0}(\theta)+f_{i2}(\theta)(r-r_h)^2+O(r-r_h)^3$, 
$w(r,\theta)=\Omega_H+w_{2}(\theta)(r-r_h)^2+O(r-r_h)^3$ 
(where the functions $f_{ik}(\theta),w_{2}(\theta)$ are
solutions of a set 
of nonlinear second order ordinary differential equations
and $f_{00}(\theta)=0$), 
which leads to an
event horizon metric 
\begin{eqnarray}
\label{eh-m}
d\Sigma^2=h_{ij}dx^idx^j =
 f_{10}(\theta)  r_h^2 d\theta^2
+f_{20}( \theta)d\varphi_1^2
+f_{30}(\theta) d\varphi_2^2.
\end{eqnarray}
For any horizon topology,
the Hawking temperature,
 horizon area and
 the entropy of a EGB
 solution 
read 
\begin{eqnarray}
\label{S} 
\nonumber
T_H= \frac{1}{2\pi}\sqrt{\frac{f_{02}}{f_{10}}},~
A_H=4\pi^2 r_h
\int_{0}^{\pi/2}
d\theta
\sqrt{ f_{10} f_{20}f_{30}},~
S=\pi^2 r_h
\int_{0}^{\pi/2}
d\theta
\sqrt{ f_{10} f_{20}f_{30}}
\left(1+ 2\alpha  R_{\Sigma}\right),~{~}
\end{eqnarray}
with
\begin{eqnarray}
\label{S2}
\nonumber
R_{\Sigma}=\frac{1}{2r_h^2 f_{10}}
\left(
\bigg(
\frac{f_{20,\theta}}{f_{20}}
+\frac{f_{30,\theta}}{f_{30}}
\bigg)
\frac{f_{10,\theta}}{f_{10}}
 +
\frac{ f_{20,\theta}^2}{f_{20}^2}
+\frac{f_{30,\theta}^2}{f_{30}^2}
-\frac{ f_{20,\theta}f_{30,\theta}}{f_{20}f_{30}}
-\frac{2f_{20,\theta\theta} }{f_{20}}
-\frac{2f_{30,\theta\theta} }{f_{30}}
\right).
\end{eqnarray} 
BHs and BRs are distinguished by the boundary conditions they satisfy at
$\theta=0$. 
For BRs one imposes 
$f_3=\partial_\theta f_0=\partial_\theta f_1=\partial_\theta f_2=\partial_\theta w=0$
for $r_h\leq r< r_b$,
 and $f_2=\partial_\theta f_0=\partial_\theta f_1=\partial_\theta f_3=\partial_\theta w=0$ for $r\geq r_b$
 (with $r_b>r_h$ 
an input parameter
roughly corresponding to the ring's $S^1$ radius 
\cite{Kleihaus:2009dm,Kleihaus:2009wh}).
The generalizations of the MP BHs have 
$f_2=\partial_\theta f_0=\partial_\theta f_1=\partial_\theta f_3=\partial_\theta w=0$
  for any $r\geq r_h$. 
For both types of solutions, the conditions satisfied by the metric functions at
$\theta=\pi/2$ are 
$f_3=\partial_\theta f_0=\partial_\theta f_1=\partial_\theta f_2=\partial_\theta w=0$.
 
 As $r\to \infty$, the Minkowski spacetime background (\ref{mink}) is recovered,
 with 
$f_0=f_1= 1$, 
$f_2=r^2\sin^2 \theta$, 
$f_3=r^2\cos^2 \theta$,  
$w=0$.
  The mass $M$ and the angular momentum $J$ of solutions are read from the asymptotic expansion 
   of the metric functions,
 $f_0=1- {8 M}/{3 \pi r^2}+\dots$ ,  $w=  {4J}/{\pi r^4}+\dots$ .

 A crucial ingredient of our approach is to use a set of 
 background functions which take automatically into account the sets
 of boundary conditions on the boundaries that determine the topology of the horizon.
 One defines $f_i=F_i f_i^{(b)}$ and $w=F_4 +w^{(b)}$, where
  $f_i^{(b)}$ and $w^{(b)}$ are the functions of the 
  corresponding solution in  Einstein gravity\footnote{
Both the MP BH
and the Emparan-Reall balanced BR can be written in the coordinate
system (\ref{metric}), with a complicated expression of the metric functions 
\cite{Kleihaus:2014pha}.
}.
It follows that the boundary conditions satisfied by $F_i$ are $\partial_r F_i=0$ at the horizon,
$F_i=1$ ($i=0,\dots,3$), $F_4=0$ at infinity and 
$\partial_\theta F_i=0$ on the symmetry axes ($\theta=0,\pi/2$).
We then employ a numerical
scheme developed in \cite{Kleihaus:2009dm,Kleihaus:2009wh} which uses a  Newton-Raphson method
to solve for the $F_i$, whilst ensuring that all the EGB equations are satisfied. 
 Mapping spatial infinity to the finite value 
$\bar r=1$ via $\bar r=1-r_h/r$,
 the numerical errors for the functions are estimated to be of order of $10^{-3}$.
The reader is referred to the Appendix B of 
\cite{Kleihaus:2009dm} for details of the procedure.

\subsection{The solutions }

Detailed discussions of the properties of the 
MP BH and BR solutions in Einstein gravity 
have appeared in various places
in the literature, see $e.g.$ the review work
 \cite{Emparan:2006mm}. 
Here we shall briefly mention only some
features which occur later when discussing the numerical EGB generalizations.
For a given mass, the MP  BHs
describe a one parameter family of solutions
which interpolate between the static BHs and a maximally rotating configuration
which is singular,
with $j=1$ and zero temperature and  horizon area\footnote{The MP solution
with rotation in a single plane has 
$a_H=t_H= \sqrt{1-j^2}$.
For the corresponding BRs 
one finds instead
$j=(1+x^2)^3/(4x(1+x^4))$,
$a_H=x(x^2-1)/(1+x^4)$
and
$t_H=(x^2-1)/(2x)$,
with $x\geq 1$.
}.
The picture for BRs is more complicated, with the existence of two branches 
of solutions
which branch off from a cusp at 
$j=j_{(min)}\simeq 0.918$,
$a_H=a_{H(max)} \simeq 0.354$
and 
$t_H \simeq 0.707$.
One of these branches corresponds to  thick BRs and  has
a small extent, meeting at $(j, aH) = (1, 0)$ the singular MP solution.
No upper bound on
$j$
exists for the thin BRs branch,
which at large angular momentum 
 effectively become boosted black strings. 
 Moreover, in the region $j_{(min)} < j <1$   three black objects
with the same mass and angular momentum coexist,
thus violating BH uniqueness.

Starting from the respective solutions in Einstein gravity 
($F_0=F_1=F_2=F_3=1$, $F_4=0$), 
we have generated branches of BHs and BRs 
by increasing the GB coupling constant $\alpha$ 
from zero, while keeping the parameters $r_h,w_h$
(and $r_b$ for BRs) fixed.
To assure that the solutions are regular,
we have monitored a number of invariant
quantities such as the Ricci and Kretschmann scalars.
All solutions we have found are finite in the full domain of integration,
in particular at $r=r_h$ and at $\theta=0,\pi/2$.
Also, let us mention that,
as with the Einstein gravity case,
	the generic BRs describe 
 $unbalanced$ configurations 
(which thus would possess one extra parameter). 
As such, for given $(r_h,r_b)$, solutions
 without conical singularities are found for
a single value of $w_h$ only. All 
 BRs reported in this work  are balanced BRs.
 
\medskip

For the case of BHs with a spherical event horizon topology,
the results of the numerical investigation are shown in 
Fig.~\ref{fig1} (right panel),
as  resulting from several hundreds of data points.
Let  us mention that, 
different from the previous Section,
this covers only partially the full
domain of existence of the solutions.
In particular, we could not construct accurate enough rotating solutions
emerging from static solutions close to the singular one with $x=1$,
which we think is only a numerical issue. 
We also remark that
 all generalizations of the MP BHs we have investigated
satisfy the  $j<1$ bound, 
which is likely to hold in the presence of a GB term.

 As with the BHs with two equal angular momenta, 
two boundaries of the domain displayed 
in Fig.~\ref{fig1} (right panels)
are provided by:
$i)$  the (Einstein gravity) MP solutions
and
 $ii)$ the static (spherically symmetric) BHs discussed in Section 2.
In addition, there is also $iii)$ the set of critical solutions,
which is approached for a maximal value of the GB coupling constant.
Unfortunately,
due to severe numerical difficulties,
 we could not  clarify the  
meaning of  this critical set.
There the numerics fails to converge,
without an obvious pathological behaviour of the solutions
in its vicinity (see also the comments at the end of this Section).  
However, 
 it is tempting to conjecture that this set
emerges
at the critical (singular) MP solution,
and ends  in the static singular solution 
in the EGB model 
(two regions which were not possible to investigate numerically).

\begin{figure}[h!]
\begin{center}
\includegraphics[height=.26\textheight, angle =0]{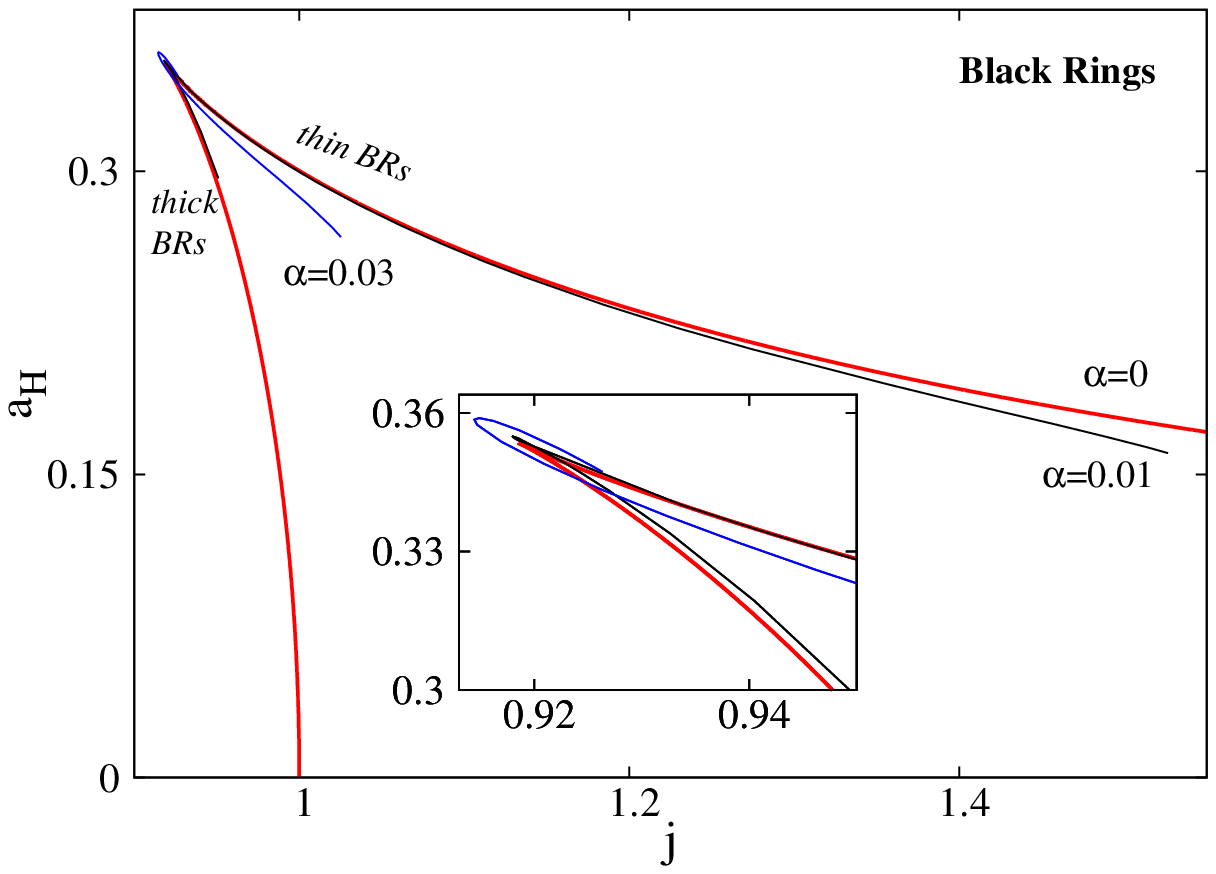} 
\includegraphics[height=.26\textheight, angle =0]{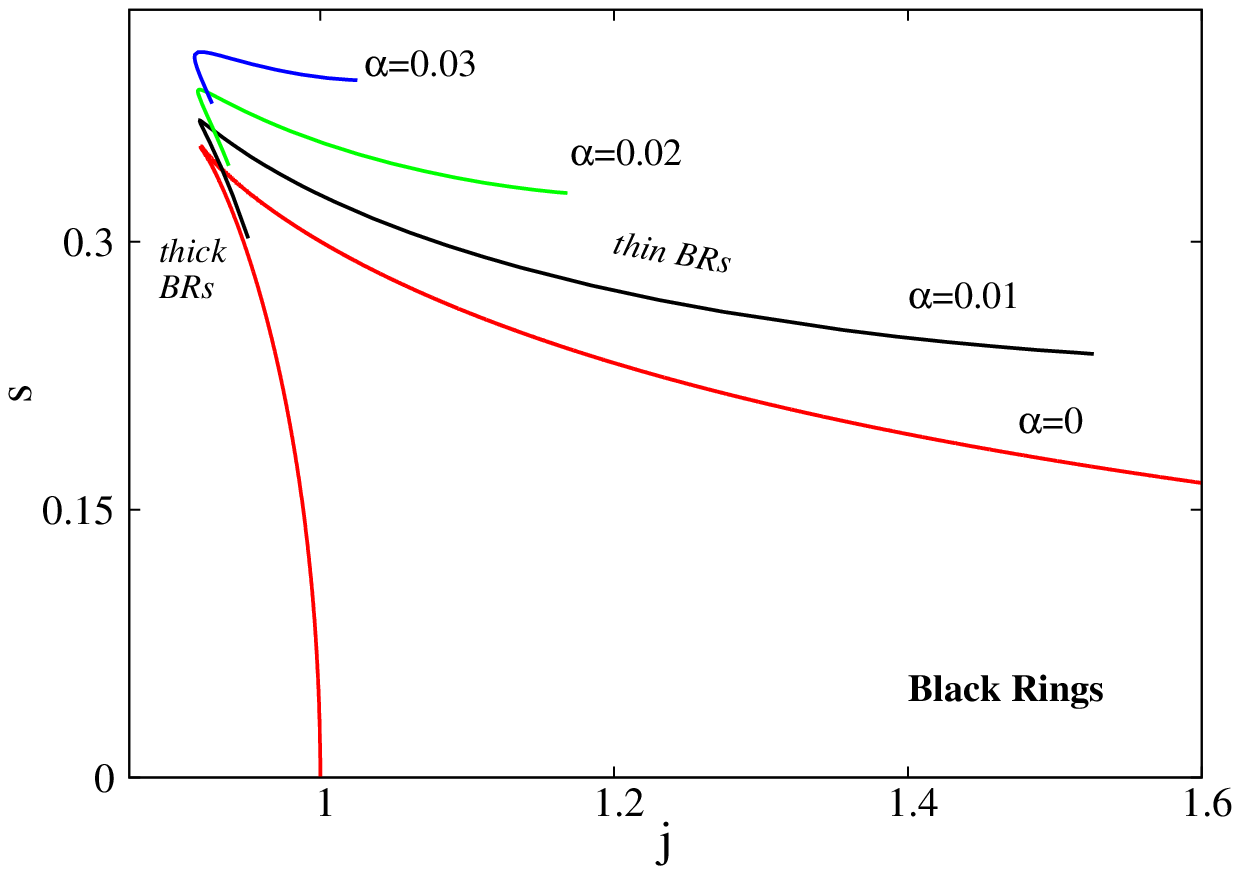} 
\end{center}
\caption{The reduced horizon area  $a_H$ 
and entropy $s$
are shown as a function of the reduced angular momentum $j$
for balanced black ring solutions in Einstein-Gauss-Bonnet theory
with several values of the GB parameter $\alpha$.  }
\label{fig2}
\end{figure}
 
\medskip

We have also managed to construct
EGB generalizations of the 
Emparan-Reall (balanced) BRs, 
several results being shown in Fig.~\ref{fig2}.
The numerical investigation was less systematic in this case,
and we did not aim to scan their domain of existence. 
We only remark that
as the GB term is added,
the two branches of BRs mentioned above persist for small values
of $\alpha$ (together with the corresponding BHs with an $S^3$ topology of the horizon). 
Thus non-uniqueness \textit{persists} in EGB theory.
We also  notice\footnote{One remarks that
 the horizon area of BRs, when considered as a function of angular momentum (at fixed mass), 
 exhibits  a "loop" 
 in the vicinity of $j_{min}$  
 (instead of a spike, as for $\alpha=0$),
see the inset in Fig.~\ref{fig2}.
The existence of such loops in
the phase diagram of spinning solutions has also been noticed 
in some $d=4$ models with non-Abelian matter fields 
\cite{Kleihaus:2005fs}. 
}
 the existence of BR solutions
violating the GR bounds, $i.e.$ with 
$j<j_{(min)}$ 
and 
$a_H>a_{H(max)}$, see Fig.~\ref{fig2}. 
However, rather unexpectedly, in our calculations
both the BH branch and the thick BR branch terminate 
before an extremal singular configuration with vanishing area is reached,
and also the thin BR branch cannot be extended to (arbitrarily)
large values of $j$,  see Fig.~\ref{fig2} (right panel).

We therefore conjecture that as $\alpha$ increases,
the domain of existence of all
three branches of black objects decreases.
Consequently, the region where BHs and BRs coexist 
also decreases with $\alpha$.
As such, beyond a first critical value of $\alpha$,
 BHs and BRs no longer coexist, 
while beyond a second critical value
only BHs persist.

The conclusion that (balanced) rotating BRs exist only
up to a maximal value of the GB coupling 
should not come as a surprise, though,
since such a behaviour was already found
for static BRs in EGB theory
\cite{Kleihaus:2009dm}. 
There the existence of a maximal  
$\alpha$ follows from conditions on the
metric functions 
for a regular horizon \cite{Kleihaus:2009dm},
being
analogous to that found in  the black string case
\cite{Kobayashi:2004hq}.
For balanced thin BRs, 
our numerical results indicate that, indeed,
a similar condition should hold
and thus impose a maximal value for $\alpha$.
For BHs and balanced thick BRs, however, a different 
condition should
impose a maximal value of $\alpha$ 
and limit their domain of existence. 
While we have not been able to clarify its origin,
and simply noticed its presence,
we conjecture that 
this could be explained
by investigating 
the expressions of higher order terms 
in the near horizon expansion of the solutions\footnote{
We mention that the four dimensional BHs in
EGB-dilaton theory  
\cite{Kleihaus:2011tg}
also possess a set of critical solutions 
where the numerics stop  to converge.
However, in that case it was possible to explain this feature  
by a study of the second order terms in the near horizon expansion
of the solutions.
}.

\section{Further remarks}

The main purpose of this paper was to present
a preliminary discussion of three different classes of 
rotating Black Holes (BHs) in $d=5$ EGB theory.
These are the generalizations	
of the Myers-Perry (MP) BHs with one and  two (equal-magnitude) angular momenta,
and of the Emparan-Reall balanced Black Rings (BRs).
 
The results here strongly suggest that, as expected, any Einstein gravity solution
 possesses generalizations with a GB term.
Also, the upper bound (for a given mass) on the value of 
the GB coupling constant $\alpha$ found in the static case
 holds as well for rotating solutions.
Moreover, the solutions with a spherical horizon topology
still satisfy the GR bound on the angular momentum, $j\leq 1$.

For the case of doubly spinning BHs,
 the inclusion of a GB term in the action
does not affect most of the qualitative features of the known MP solutions.
 This holds as well for  the
extremal EGB BHs,
which are reported here for the first time 
in the literature.
Although 
our results for the singly spinning black objects are only partial,
they indicate the existence of a different
type of critical behaviour of the solutions  at maximal $\alpha$,
which we could  not  yet clarify.
Nevertheless, we have found that 
the non-uniqueness of solutions
(with the existence of three different objects
with the same mass and angular momentum)
holds in EGB theory for small enough
$\alpha$ only. 
Further progress in the study of EGB  solutions
with a single $J$
seems to require a different numerical scheme.

Finally, it would be interesting to compare the results in this work
with those found 
in \cite{Konoplya:2020fbx}
within a perturbative approach.

\section*{Acknowledgements}
B.K. and J.K. gratefully acknowledge support 
by the DFG Research Training Group 1620 \textit{Models of Gravity} and DFG project Ku612/18-1.
E.R. gratefully acknowledges the support of the Alexander von Humboldt Foundation,
and also the support  from the projects CERN/FIS-PAR/0027/2019,
 PTDC/FIS-AST/3041/2020,  
CERN/FIS-PAR/0024/2021 and 2022.04560.PTDC.  
His work was also
supported  by the  Center for Research and Development in Mathematics and Applications (CIDMA) through the Portuguese Foundation for Science and Technology (FCT -- Fundac\~ao para a Ci\^encia e a Tecnologia), references  UIDB/04106/2020 and UIDP/04106/2020.  
This work has further been supported by  the  European  Union's  Horizon  2020  research  and  innovation  (RISE) programme H2020-MSCA-RISE-2017 Grant No.~FunFiCO-777740 and by the European Horizon Europe staff exchange (SE) programme HORIZON-MSCA-2021-SE-01 Grant No.~NewFunFiCO-101086251.


	

\begin{thebibliography}{99}
\bibitem{Lovelock:1971yv}
  D.~Lovelock,
  J.\ Math.\ Phys.\  {\bf 12} (1971) 498.
\bibitem{Antoniadis:1997eg}
  I.~Antoniadis, S.~Ferrara, R.~Minasian and K.~S.~Narain,
  Nucl.\ Phys.\  B {\bf 507} (1997) 571
  [arXiv:hep-th/9707013];
\\
  S.~Ferrara, R.~R.~Khuri and R.~Minasian,
  Phys.\ Lett.\  B {\bf 375} (1996) 81
  [arXiv:hep-th/9602102].
\bibitem{1} 
  D.~J.~Gross and E.~Witten,
  Nucl.\ Phys.\  B {\bf 277} (1986) 1;
 \\
  R.~R.~Metsaev and A.~A.~Tseytlin,
  Phys.\ Lett.\  B {\bf 191} (1987) 354;
 \\
   C. G. Callan, R. C. Myers, and M. J. Perry, 
   Nucl. Phys. {\bf B311} (1988) 673.  
\bibitem{Myers:1987yn}
  R.~C.~Myers,
  Phys.\ Rev.\  D {\bf 36} (1987) 392.
\bibitem{Deser}
D.~G.~Boulware and S.~Deser, 
Phys.\ Rev.\ Lett.\  {\bf 55} (1985) 2656;
\\
  J.~T.~Wheeler,
  Nucl.\ Phys.\  B {\bf 268} (1986) 737.
\bibitem{Tangherlini:1963bw}
  F.~R.~Tangherlini,
  Nuovo Cim.\  {\bf 27} (1963) 636.
\bibitem{Jacobson:1993xs}
  T.~Jacobson and R.~C.~Myers,
  Phys.\ Rev.\ Lett.\  {\bf 70} (1993) 3684
  [arXiv:hep-th/9305016].
 \bibitem{Wald:1993nt}
  R.~M.~Wald,
  Phys.\ Rev.\  D {\bf 48} (1993) 3427
  [arXiv:gr-qc/9307038].
	
\bibitem{Garraffo:2008hu}
  C.~Garraffo and G.~Giribet,
  Mod.\ Phys.\ Lett.\  A {\bf 23} (2008) 1801
  [arXiv:0805.3575 [gr-qc]].
 \bibitem{Charmousis:2008kc}
  C.~Charmousis,
  Lect.\ Notes Phys.\  {\bf 769} (2009) 299
  [arXiv:0805.0568 [gr-qc]].
\bibitem{Anabalon:2009kq}
  A.~Anabalon, N.~Deruelle, Y.~Morisawa, J.~Oliva, M.~Sasaki, D.~Tempo and R.~Troncoso,
  Class.\ Quant.\ Grav.\  {\bf 26} (2009) 065002
  [arXiv:0812.3194 [hep-th]].
\bibitem{Kim:2007iw}
  H.~C.~Kim and R.~G.~Cai,
  Phys.\ Rev.\  D {\bf 77} (2008) 024045
  [arXiv:0711.0885 [hep-th]];
  \\ 
  S.~Alexeyev, N.~Popov, M.~Startseva, A.~Barrau and J.~Grain,
  J.\ Exp.\ Theor.\ Phys.\  {\bf 106} (2008) 709
  [arXiv:0712.3546 [gr-qc]].
\bibitem{Konoplya:2020fbx}
R.~A.~Konoplya and A.~Zhidenko,
Phys. Rev. D \textbf{102} (2020) no.8, 084030
[arXiv:2007.10116 [gr-qc]].
 \bibitem{Brihaye:2008kh}
  Y.~Brihaye and E.~Radu,
  Phys.\ Lett.\  B {\bf 661} (2008) 167
  [arXiv:0801.1021 [hep-th]].
\bibitem{Brihaye:2010wx}
Y.~Brihaye, B.~Kleihaus, J.~Kunz and E.~Radu,
JHEP \textbf{11} (2010), 098
[arXiv:1010.0860 [hep-th]].
 
\bibitem{Emparan:2001wn}
  R.~Emparan and H.~S.~Reall,
  Phys.\ Rev.\ Lett.\  {\bf 88} (2002) 101101
  [arXiv:hep-th/0110260].
\bibitem{Myers:1986un}
  R.~C.~Myers and M.~J.~Perry,
  Annals Phys.\  {\bf 172} (1986) 304.
	
	
\bibitem{Kleihaus:2009dm}
B.~Kleihaus, J.~Kunz and E.~Radu,
JHEP \textbf{02} (2010), 092
[arXiv:0912.1725 [gr-qc]].

\bibitem{Kleihaus:2009wh}
  B.~Kleihaus, J.~Kunz and E.~Radu,
  Phys.\ Lett.\  B {\bf 678} (2009) 301
  [arXiv:0904.2723 [hep-th]].
 
\bibitem{Kleihaus:2014pha}
B.~Kleihaus, J.~Kunz and E.~Radu,
JHEP \textbf{01} (2015), 117
[arXiv:1410.0581 [gr-qc]].

\bibitem{Emparan:2006mm}
R.~Emparan and H.~S.~Reall,
Class. Quant. Grav. \textbf{23} (2006), R169
[arXiv:hep-th/0608012 [hep-th]].

\bibitem{Kleihaus:2005fs}
B.~Kleihaus, J.~Kunz and U.~Neemann,
Phys. Lett. B \textbf{623} (2005), 171-178
[arXiv:gr-qc/0507047 [gr-qc]];
\\
B.~Kleihaus, J.~Kunz, F.~Navarro-Lerida and U.~Neemann,
Gen. Rel. Grav. \textbf{40} (2008), 1279-1310
[arXiv:0705.1511 [gr-qc]].


\bibitem{Kobayashi:2004hq}
  T.~Kobayashi and T.~Tanaka,
  Phys.\ Rev.\  D {\bf 71} (2005) 084005
  [arXiv:gr-qc/0412139].
\bibitem{Kleihaus:2011tg}
B.~Kleihaus, J.~Kunz and E.~Radu,
Phys. Rev. Lett. \textbf{106} (2011), 151104
[arXiv:1101.2868 [gr-qc]];
\\
B.~Kleihaus, J.~Kunz, S.~Mojica and E.~Radu,
Phys. Rev. D \textbf{93} (2016) no.4, 044047
[arXiv:1511.05513 [gr-qc]].
	
\end{thebibliography}
\end{document}